\documentstyle[prl,aps,amssymb,multicol,epsf]{revtex}

\title{Radial distribution function of semiflexible polymers}

\author{Jan Wilhelm and Erwin Frey} 

\address{Institut f\"ur Theoretische Physik, Technische Universit\"at
  M\"unchen, 85747 Garching, Germany}

\date{Phys. Rev. Lett. 77, 2581 (1996)}

\begin{document}

\maketitle

\begin{abstract}
  We calculate the distribution function of the end--to--end distance of a
  semiflexible polymer with large bending rigidity. This quantity is directly
  observable in experiments on single semiflexible polymers (e.g., DNA, actin)
  and relevant to their interpretation. It is also an important starting point
  for analyzing the behavior of more complex systems such as networks and
  solutions of semiflexible polymers.  To estimate the validity of the obtained
  analytical expressions, we also determine the distribution function
  numerically using Monte Carlo simulation and find good quantitative
  agreement.
\end{abstract}


\begin{multicols}{2}

\narrowtext

  While we have a comparably complete theoretical picture of highly flexible
  chain molecules, the statistical mechanics of semiflexible polymers is a
  field with a number of open questions that has received renewed attention
  lately. Considerable motivation stems from the crucial importance of the
  elasticity of biopolymers like spectrin, actin, and microtubules for the
  mechanical properties of cells~\cite{sackmann:94}. Recent advances in
  visualizing and manipulating such macromolecules have provided unique
  experimental tools for the study of the static and dynamic properties of
  single filaments \cite{nagashima-asakura:80,smith-finzi-bustamante:92,%
    ott-magnasco-simon-libchaber:93,kaes-etal:93,gittes-etal:93}. A
  quantitative measure for a polymer's flexibility is its persistence length
  $\ell_p$, which is the characteristic length governing the decay of
  tangent--tangent correlations.  Nature provides polymers of very different
  stiffness, e.g., $\ell_p \approx 10 \, \text{nm}$ for
  spectrin~\cite{svoboda-etal:92}, $\ell_p \approx 50 \, \text{nm}$ for
  DNA~\cite{marko-siggia:95}, $\ell_p \approx 17 \, \text{$\mu$m}$ for
  actin~\cite{ott-magnasco-simon-libchaber:93,gittes-etal:93} and $\ell_p
  \approx 5.2 \, \text{mm}$ for microtubules~\cite{gittes-etal:93}. On length
  scales larger than a few $\ell_p$ a polymer with contour length $L \gg
  \ell_p$ (flexible polymer) can be described as a self--avoiding freely
  jointed chain. For molecules with $t := L/\ell_p$ of the order of one
  (semiflexible polymer) this is not possible, as can be seen immediately by
  comparing a typical contour (e.g., from Ref.~\cite{kaes-etal:93}) to the
  random walk corresponding to a freely jointed chain. Even for long strands of
  DNA with $t \approx 1000$ measurements of the molecule's extension at large
  forces have revealed significant deviations from the freely jointed chain
  model \cite{smith-finzi-bustamante:92}, which find their explanation in
  bending elasticity \cite{marko-siggia:95}. Thus it is essential to consider
  models which take chain rigidity into account. For sufficiently stiff
  chains effects resulting from self--avoidance can be neglected due to the
  strong energetic suppression of configurations where the chain folds back
  onto itself.  Furthermore the polymers under consideration can be regarded as
  inextensible \cite{note:1}. The corresponding model is the wormlike chain
  introduced by Kratky and Porod almost 50 years ago \cite{kratky-porod:49}.
    
  A central quantity for characterizing the conformations of single polymer
  chains is the distribution function $G({\bf r};L)$ of the end--to--end
  distance ${\bf r}$ for given contour length $L$ and persistence length
  $\ell_p$. For models like the wormlike chain with only short-range
  interactions between monomers, $G({\bf r}; L)$ actually is the probability
  density of finding any two monomers at relative position ${\bf r} = {\bf
    r}(s) - {\bf r}(s')$ where $L = |s-s'|$ is the distance between the
  monomers along the chain.  For a freely jointed phantom chain $G({\bf r};L)$
  is known exactly~\cite{yamakawa} and for many purposes is well approximated
  by a simple Gaussian.  Complications arise when the self--avoidance of real
  chains is taken into account \cite{descloizeaux:90}.  In previous
  investigations of the wormlike chain $G({\bf r}; L)$ was obtained
  approximately for almost fully flexible polymers (large $t$) in the form of
  corrections to the Gaussian distribution function up to order $t^{-2}$
  \cite{daniels:52,gobush-yamakawa-stockmayer-magee:72}.  For general $t$ only
  the lowest three even moments were calculated analytically
  \cite{hermans-ullman:52}. Higher even moments were obtained by numerical
  techniques~\cite{yamakawa-fujii:74}.  For polymers close to the stiff limit
  all even moments of the distribution function were calculated in an expansion
  in $t$ \cite{hermans-ullman:52,norisuye-murakama-fujita:78}, but the
  expressions obtained for $G({\bf r}; L)$ in this limit are only given up to
  quadratures \cite{norisuye-murakama-fujita:78,yamakawa-fujii:73a} and do not
  show the correct qualitative behavior when integrated numerically.
  
  In this letter we determine $G({\bf r}; L)$ in two- and three-dimensional
  embedding space in an approximation valid for small $t$ and compare
  the analytical expressions obtained to data from a Monte Carlo simulation.
  The range of validity of our results almost extends to values of the bending
  rigidity where the Daniels
  approximation~\cite{daniels:52,gobush-yamakawa-stockmayer-magee:72} becomes
  applicable. 

  For our analytical calculations we adopt a continuum version of the wormlike
  chain model where the polymer is represented by a differentiable space curve
  ${\bf r} (s)$ of length $L$ parameterized to arc length
  \cite{saito-takahashi-yunoki:67}. Its statistical properties are determined
  by an effective free energy
  \begin{equation}
    \label{worham} 
    {\cal H} = 
    \frac{\kappa}{2} \int_0^L\!\!ds
    \left[ \frac{\partial {\bf t}(s)}{\partial s}\right]^{\!\!2} \, ,
  \end{equation}
  where ${\bf t}(s) = \partial {\bf r} (s) / \partial s$ is the tangent vector
  at arc length $s$. The resulting persistence lengths are $\ell_p = \kappa/k_B
  T$ for $d=3$ and $\ell_p = 2\kappa/k_B T$ for $d=2$, where $d$ is the
  dimension of the embedding space \cite{note:4}. The inextensibility of the
  chain is expressed by the local constraint $|{\bf t} (s)|=1$ which leads to
  non-Gaussian path integrals. Note that this local curvature model is
  equivalent to a one-dimensional nonlinear $\sigma$-model. While a few
  quantities like $\langle R^2 \rangle$ and $\langle R^4 \rangle$ can be
  obtained exactly, one has to resort to some approximative scheme to calculate
  the end--to--end distribution function
  \begin{equation}
    \label{eqn:def_of_G}
    G({\bf r}; L) = \langle \delta( {\bf r} - {\bf R}) \rangle \, ,
  \end{equation}
  where ${\bf R} := {\bf r} (L) - {\bf r} (0)$ and $\delta({\bf r})$ is the
  Dirac $\delta$-function.
  
  In order to understand the effect of possible approximations to
  Eq.~(\ref{eqn:def_of_G}) for stiff polymers, it is instructive to
  recapitulate the classical problem of bending a rigid rod. The energy of a
  straight rod of length $L$ and bending modulus $\kappa$ is an almost linear
  function of its end-to-end distance $r$: $E_{\text{cl}} \approx f_c(L-r)$
  where $f_c = \kappa \pi^2/L^2$ is the critical force for the onset of the
  Euler instability (since the radial distribution function does not specify
  the direction of the tangent vectors at $s=0$ and $s=L$, the appropriate
  boundary conditions are open ends). Neglecting fluctuations around the
  classical contour this would lead to an end--to--end distribution function
  with maximum weight at $r=L$, $G({\bf r};L) \propto \exp [- f_c (L-r)/(k_B
  T)]$. But, for a completely stretched chain there is up to global rotations
  only one possible configuration and consequently the end--to--end
  distribution function has to vanish at full extension. Hence it is essential
  to take into account entropy effects.  While stiff chains are energy
  dominated, more flexible chains are mostly governed by the entropic effects
  with the limit being the freely jointed chain.  Approaching full extension,
  however, $G({\bf r}; L)$ must vanish for all chains. This shows that while
  $G({\bf r}; L)$ tends to the distribution $\delta(L-r)/4\pi L^2$ ($d=3$) for
  $\kappa \to \infty$ and $t \to 0$, it can never be expanded in $t$ around
  this limit (consider the probability of full extension). For this reason, it
  is impossible to obtain the distribution function from an expansion of all
  even moments in terms of $t$ \cite{norisuye-murakama-fujita:78}. Softening
  the constraint of fixed contour length affects both energetic and entropic
  contributions to the distribution function in essential ways. If it is
  relaxed to the point of fixing only $\langle \int_0^L ds {\bf t}^2(s)
  \rangle$ by means of a single Lagrange multiplier (e.g.,
  Ref.~\cite{freed:71}), the distribution functions obtained are essentially
  Gaussian and will not show the correct qualitative behavior for stiff
  polymers. Failure to reproduce the vanishing of $G({\bf r}; L)$ near full
  extension also results when approximations are used which neglect essential
  parts of the fluctuational contributions as in
  Ref.~\cite{yamakawa-fujii:73a}.
  
  For end--to--end distances $r$ close enough to full stretching and/or large
  values of $\kappa$, the typical configuration of the chain will be close to a
  straight rod. Thus the deviations of the tangent vectors from the average
  direction can be treated as small variables. For $d=3$ we parameterize the
  contour through the tangent field: ${\bf t} (s) = (a_x (s) , a_y (s) , 1) /
  \sqrt{1+a_x^2 (s) + a_y^2 (s)}$, which properly takes into account the
  constraint of inextensibility. We employ a harmonic approximation and keep
  only terms up to second order in ${\cal H}$, the measure factor, and the
  arguments of the $\delta$--function in Eq.~(\ref{eqn:def_of_G}). The error
  caused by this approximation vanishes near full extension. From here on we
  shall measure all lengths in units of $L$ and all energies in units of $k_B
  T$. With this convention we have (for $d=3$) $\kappa = \ell_p$ and $t =
  \kappa^{-1}$. We also drop the second argument in $G({\bf r}; 1)$ and make
  rotation invariance explicit by writing $G(r)$. Use of $2\pi\delta(x) = \int
  dq \exp(iqx)$, expansion of $a_x(s)$ and $a_y(s)$ in cosine series as
  appropriate for the boundary conditions of open ends, and evaluation of the
  resulting Gaussian integrals leads to
  \begin{equation}
    \label{eqn:3d_G(r)_form2}
    G(r) = \frac{2 \kappa}{4\pi {\cal N}} 
    \sum_{k=1}^\infty \pi^2 k^2  (-1)^{k+1} 
    e^{- \kappa \pi^2 k^2 (1-r)} \, ,
  \end{equation}
  where ${\cal N}$ is a normalization factor compensating the failure of the
  approximation used to conserve the normalization of $G(r)$ for finite
  $\kappa$. Details of the calculation can be found in a forthcoming
  publication \cite{wilhelm-frey:96b}. For $\kappa (1-r) \gtrapprox 0.2$ the
  distribution function is dominated by the $k=1$ term which is just our
  heuristic result from above (``Euler instability'') if the dimensional
  factors are put back in.  For $r \to 1$, however, more and more terms have to
  be considered. By writing Eq.~(\ref{eqn:3d_G(r)_form2}) as an integral over
  Fourier expanded $\delta$-functions it is possible to transform
  Eq.~(\ref{eqn:3d_G(r)_form2}) to
  \begin{eqnarray}
    \label{eqn:3d_G(r)_form3}
    G(r) = &&\frac{1}{4 \pi {\cal N}} \frac{\kappa}{2\sqrt{\pi}} 
    \sum_{\ell=1}^\infty
    \frac{1}{[\kappa(1-r)]^{3/2}} \nonumber \\
    &&\times \exp\left[-\frac{(\ell-1/2)^2}{\kappa (1-r)}\right] 
    H_2 \left[\frac{\ell-1/2}
      {\sqrt{\kappa (1-r)}}\right] \, ,
  \end{eqnarray}
  where $H_2(x) = 4 x^2 - 2$ is the second Hermite polynomial. This series
  converges very quickly for $\kappa (1-r) \lessapprox 0.2$ where the behavior
  of $G(r)$ is completely dominated by the $\ell=1$ term.
  
  In recent experiments as well as in simulations the polymer was effectively
  restricted to a $d=2$ embedding space
  \cite{ott-magnasco-simon-libchaber:93,kaes-etal:93,gittes-etal:93,%
    hendricks-etal:95}. In this case it is convenient to choose a
  parameterization ${\bf t}(s) = (\cos \phi, \sin \phi)$. The resulting
  effective free energy is quadratic in $\phi$, ${\cal H} = (\kappa/2) \int ds
  \left( {\partial \phi / \partial s} \right)^2$. In order to calculate $G(r)$
  we again approximate the constraints given by the $\delta$--functions in
  Eq.~(\ref{eqn:def_of_G}) by $0 = r_y(1) \approx \int ds \phi$ and $r = r_x(1)
  \approx 1 - \frac{1}{2} \int ds \phi^2$.  A calculation along similar lines
  as before \cite{wilhelm-frey:96b} leads to the result
  \begin{eqnarray}
    \label{eqn:2d_G(r)_form2}
    \lefteqn{G(r) = 
    \frac{1}{2\pi {\cal N}} \frac{2 \kappa}{\sqrt{\pi}} 
    \sum_{\ell=0}^\infty \frac{(2\ell-1)!!}{2^\ell \ell!} 
    \frac{1}{[2\kappa(1-r)]^{5/4}}}
    \nonumber \\
    &&\times 
    \exp\left[-\frac{(\ell+1/4)^2}{2\kappa(1-r)}\right] D_{3/2}\left[2
      \frac{\ell+1/4}{\sqrt{2\kappa(1-r)}} \right],
  \end{eqnarray}
  with $D_{3/2}(x)$ a parabolic cylinder function. The convergence
  properties of Eq.~(\ref{eqn:2d_G(r)_form2}) are similar to those of
  Eq.~(\ref{eqn:3d_G(r)_form3}). 
  
  In order to assess the quality of our approximations, we have used Monte
  Carlo simulation to evaluate $G(r)$ numerically. We adopted the following
  discretized version of the wormlike chain: The polymer is described as a
  chain composed of $N$ tethers of fixed length $a= 1/N$ and direction ${\bf
    t}$ with a bending energy ${\cal H}_b = \tilde\kappa a^{-1}
  \sum_{i=1}^{N-1} {\bf t}_i \cdot {\bf t}_{i+1}$ \cite{note:2}. The standard
  Metropolis algorithm was used to measure $G(r)$. We found that results cease
  to depend appreciably on $N$ as soon as there are three to four segments in
  one persistence length. On the order of $10^6$ MC-steps per segment were
  performed.  Final results were obtained by averaging over several independent
  runs. The accuracy of $\langle R \rangle$ obtained was typically of the order
  of 0.5\%.  Measured expectation values $\langle R^2 \rangle$ and $\langle
  R^4\rangle$ were in agreement with known exact expressions up to the
  estimated statistical errors.
  
\begin{figure}
  \narrowtext
    \begin{center}
      \epsfbox{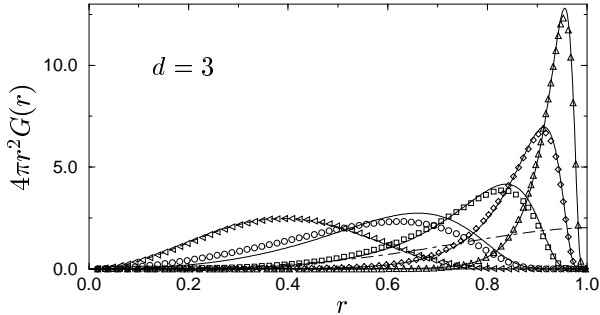}
      \caption{\label{fig:3d_fit}
        Comparison of $G(r)$ from the Monte Carlo simulation (symbols) for $t=$
        10, 5, 2, 1 and 0.5 (left to right) to Eq.~(\ref{eqn:3d_G(r)_form2})
        (solid lines). The dashed and dot-dashed lines show the second Daniels
        distribution for $t = 10$ and $t = 2$. For $t=10$ it almost coincides
        with the numerical data while it is quite far off for $t = 2$. Error
        bars for the Monte Carlo data are approximately of the size of the
        symbols.} 
    \end{center} 
  \end{figure} 

  Fig.~\ref{fig:3d_fit} shows a comparison of the normalized $G(r)$ from
  Eq.~(\ref{eqn:3d_G(r)_form2}) to the data from our Monte Carlo simulation.
  Note that there is no free parameter to adjust the curves. The curves for the
  $d=2$ case are qualitatively similar and the agreement between theory and the
  $d=2$ Monte Carlo simulation is of equal quality. It can be improved somewhat
  further by applying a simple correction procedure specific to the $d=2$ case
  \cite{wilhelm-frey:96b}. The general observation is that
  Eqs.~(\ref{eqn:3d_G(r)_form2}), (\ref{eqn:3d_G(r)_form3}) and
  (\ref{eqn:2d_G(r)_form2}) reproduce the data qualitatively right and are good
  approximations for $\ell_p \gtrapprox 0.5 L$, that is $t \lessapprox 2$.
  However, even for large values of $\kappa$ one would not expect the harmonic
  approximation to yield acceptable results for small $r$ since the curvature
  of the polymer must then be large.  The somewhat surprising quality of the
  approximation in this region can be attributed to the fact that there the
  distribution function is dominated by the energy $E_{\text{cl}}(r)$ of the
  most probable configuration. As discussed below
  Eq.~(\ref{eqn:3d_G(r)_form2}), the dominant linear term of $E_{\text{cl}}(r)$
  is reproduced by the harmonic approximation, which corresponds to the well
  known fact that $f_c$ can be obtained by looking only at infinitesimal
  compressions. For $r \approx 0$, however, the deviations of $E_{\text{cl}}$
  from the linear form get significant for all $\kappa$. While this is
  irrelevant for most applications of the distribution function due to the
  small probability of such configurations, it implies that the ring-closure
  probability $G(0)$ is not reproduced correctly by our approach.  Instead one
  should expand the configurations around the contour of minimal energy as in
  Ref.~\cite{shimada-yamakawa:84} where the resulting expressions for $G(0)$
  are evaluated numerically yielding results in agreement with our Monte Carlo
  data.
  
  Another possibility to check the validity of the obtained distribution
  functions is to compare their moments to known results. The moments $\langle
  R^n \rangle$ can either be calculated from the given expressions for $G(r)$
  or by applying the harmonic approximation directly to the generating function
  $\langle e^{fR} \rangle$.  The latter method yields an expansion of $\langle
  R^n \rangle$ in $t$ which is seen to be correct to ${\cal O} (t^2)$ when
  compared to the results of Ref.~\cite{norisuye-murakama-fujita:78}.
  Calculating the moments directly from the $\ell=1$ term of
  Eq.~(\ref{eqn:3d_G(r)_form3}) for $\kappa(1-r) < 0.2$ and the $k=1$ term of
  Eq.~(\ref{eqn:3d_G(r)_form2}) for $\kappa(1-r) > 0.2$ produces analytical
  expressions which can be expanded in terms of $t$ only up to a small
  correction vanishing like $\exp[-\text{const.}\kappa]$. This leads to
  expansion coefficients that are not rational numbers but differ very slightly
  from the results of the generating function method (by about 0.1\% for
  $t=1$). Calculation from the distribution function has the advantage that
  averages $\langle R^\alpha \rangle$ with arbitrary $\alpha > -d+1$ can be
  obtained. The results indicate that the expansion derived in
  \cite{norisuye-murakama-fujita:78} for $\alpha = 2n$ with $n=0,1,\ldots$ is
  valid for all $\alpha > -d+1$ at least up to ${\cal O}(t^2)$.
    
  A quantity of immediate experimental interest is the force-extension
  relation. For sufficiently stiff polymers it can be obtained from the given
  $G(r)$ for arbitrary forces in situations where the ends of the polymer can
  rotate freely. The well known strong-force limit (e.g.,\ 
  \cite{marko-siggia:95}) is reproduced by our distribution functions for
  arbitrary stiffness since it is determined by the behavior of $G(r)$ for $r
  \to 1$ where the harmonic approximation is expected to be good for all values
  of $\kappa$: For large forces $f$ we have $r \approx 1$ and consequently
  $G(r)$ will be dominated by the $\ell = 1$ term of
  Eq.~(\ref{eqn:3d_G(r)_form3}). The relevant free energy is thus $F = -\log
  G(r) + f(1-r) \approx 1/(4\kappa(1-r)) + f(1-r)$, where we neglected
  logarithmic terms since $1-r \to 0$ for $f \to \infty$. Since the
  corresponding distribution function will be strongly peaked for large $f$,
  $\langle r \rangle$ is given by the position of the minimum of $F$: $\langle
  r \rangle = 1 - 1/\sqrt{4 \kappa f}$ which agrees with
  Ref.~\cite{marko-siggia:95}. Force-extension relations for small forces can
  be obtained from moments of the distribution function. Our results are thus
  consistent with existing linear response treatments
  \cite{mackintosh-kaes-janmey:95,kroy-frey:96} to ${\cal O}(t^2)$. If the
  polymer's orientation at one end is fixed, the character of the response
  depends on the direction of the applied force and $G({\bf r} | {\bf u_0}; L)$
  is needed \cite{wilhelm-frey:96b} (notation of
  Ref.~\cite{yamakawa-fujii:73a}). Note that the linear response can be
  evaluated exactly for arbitrary values of $t$ in this case
  \cite{kroy-frey:96}.
  
  Using fluorescence microscopy, it is possible to visualize the thermal
  undulations of single actin filaments constrained to quasi two-dimensional
  configurations
  \cite{ott-magnasco-simon-libchaber:93,kaes-etal:93,gittes-etal:93}. Both
  contour lengths and distances in embedding space can be measured from the
  resulting images. If locality of the interactions along the chain is assumed,
  different segments of one physical polymer of length $L_0$ are statistically
  independent. One can thus probe different lengthscales by obtaining
  experimental distribution functions $G(r; L)$ for different $L \le L_0$.
  Using only $\ell_p$ as free parameter, our theoretical results can be fitted
  to the experimental distributions. The quality of the fits for different $L$
  will indicate at what scales actin can actually be described by the wormlike
  chain model.  This might reveal new interesting physics and help to clarify
  some of the ambiguities that arise in normal-mode analyses of actin
  flickering: The average squared amplitudes of the normal modes with mode
  number $k$ of the elastic Hamiltonian fail to decay like $1/k^2$ as predicted
  by the wormlike chain model
  \cite{kaes-etal:93,gittes-etal:93,hendricks-etal:95}.  Since $G(r; L)$ can be
  determined without differentiation and Fourier transformation of the observed
  contours, the radial distribution function analysis is expected to be much
  more stable against unavoidable experimental errors.  Another method that has
  been used to analyze actin flickering is the comparison of the measured decay
  of the tangent-tangent correlation function with the exact result
  $\langle{\bf t}(s){\bf t}(s')\rangle \propto \exp(-|s-s'|/\ell_p)$
  \cite{kratky-porod:49,ott-magnasco-simon-libchaber:93}.  However, since an
  exponential decay of correlations is expected for all systems with only short
  range interactions, this kind of analysis is probably not very sensitive to
  deviations from the wormlike chain model. A very interesting possibility
  would be to attach two or more markers (e.g., small fluorescent beads)
  permanently to single strands of polymers and to observe the distribution
  function of the markers' separation.  This would eliminate all the
  experimental difficulties associated with the determination of the polymer's
  contour. Note that contrary to existing methods of analysis it is not
  necessary to know the length of polymer between two markers; this quantity
  can be extracted from the observed distribution functions along with
  $\ell_p$.
  
  In conclusion, we have derived approximate analytical expressions for the
  end--to--end distribution function $G({\bf r}; L)$ of semiflexible polymers.
  The essential ingredient in our calculation is the choice of a
  parameterization for the polymer's configuration that satisfies the
  constraint of fixed contour length by construction. Comparison with Monte
  Carlo data shows good quantitative agreement for $t = L/\ell_p \lessapprox
  2$. Since the Daniels approximation is valid for $t \gtrapprox 10$, $G({\bf
    r}; L)$ of the wormlike chain is up to a crossover region now available for
  the whole range from rigid rods to random coils.  The range of stiffness
  accessible to the approximations used is highly relevant for the physics of
  rather rigid polymers like actin. Knowledge of $G({\bf r}; L)$ provides new
  possibilities for the experimental determination of $\ell_p$ as well as the
  lengthscales at which real semiflexible polymers can be described by
  the wormlike chain model.  Since the actual form of the single chain
  probability distribution function is an important input for the theory of
  many-chain systems, we hope that our work will contribute to the
  investigation of more complex questions such as the dynamics and
  viscoelasticity of networks and solutions of semiflexible polymers.
 
  We have benefited from discussions with Robijn Bruinsma, Klaus Kroy and Erich
  Sackmann. This work was supported by the Deutsche Forschungsgemeinschaft
  (DFG) under contract no. Fr~850/2 and no. SFB~266.

\end{multicols}


\begin{thebibliography}{10}
\bibitem{note:2}
To obtain the same $\ell_p$ as in the continuous model, $\tilde\kappa$ has to
  be chosen according to $\ell_p = - a/\log {\cal L}_d(\tilde\kappa/a
  k_{\text{B}} T)$, where ${\cal L}_d(x) = I_{d/2}(x)/I_{d/2-1}(x)$ and $I_n$
  are modified Bessel functions. For large $N$ this approaches the
  corresponding relation $\ell_p = \kappa/k_{\text{B}} T$ for $d=3$ and $\ell_p
  = 2\kappa/k_{\text{B}}T$ for $d=2$ of the continuum model. See e.g. M.~E.
  Fisher, Am.~J.~Phys. {\bf 32}, 343 (1964).
\end{thebibliography}
\end{document}